# A geographically distributed bio-hybrid neural network with memristive plasticity

Alexantrou Serb, Andrea Corna, Richard George, Ali Khiat, Federico Rocchi, Marco Reato, Marta Maschietto, Chirstian Mayr, Giacomo Indiveri, Stefano Vassanelli*, Themistoklis Prodromakis*[*]

**Throughout evolution the brain has mastered the art of processing real-world inputs through networks of interlinked spiking neurons. Synapses have emerged as key elements that, owing to their plasticity, are merging neuron-to-neuron signalling with memory storage and computation[1]. Electronics has made important steps in emulating neurons through neuromorphic circuits[2] and synapses with nanoscale memristors[2,3], yet novel applications that interlink them in heterogeneous bio-inspired and bio-hybrid architectures are just beginning to materialise[4,5]. The use of memristive technologies in brain-inspired architectures for computing[6,7] or for sensing spiking activity of biological neurons[8] are only recent examples, however interlinking brain and electronic neurons through plasticity-driven synaptic elements has remained so far in the realm of the imagination. Here, we demonstrate a bio-hybrid neural network (bNN) where memristors work as 'synaptors' between rat neural circuits and VLSI neurons. The two fundamental synaptors, from artificial-to-biological (ABsyn) and from biological-to-artificial (BAsyn), are interconnected over the Internet. The bNN extends across Europe, collapsing spatial boundaries existing in natural brain networks and laying the foundations of a new geographically distributed and evolving architecture: the Internet of Neuro-electronics (IoN).**

Since Hebb's intuition that neuronal connections can be strengthened by repeated activity and the following discovery of synaptic plasticity in Aplysia[9] and vertebrates[10], it has become evident that synaptic plasticity confers unique computational properties to the brain[1]. Recent findings that nanoscale memristors can emulate basic synaptic plasticity properties[11,12] have created the premise for merging nanoelectronics and living neurons to generate bio-hybrid neuronal networks (bNN) interlinked through electronics-based plasticity elements. The fields of neural interfaces, memristor-based systems and bio-inspired electronics, all independently boast demonstrations of large-scale implementations. Experimental electrophysiology is now supported by microelectrode arrays with up to 1024 recording sites in vivo[13] and >16k in vitro[14,15]. Similarly, large-scale memristive systems using >100k devices have been reported[16], as have medium-scale memristor in-situ handling platforms[17] (up to 1024 devices). Simultaneously, architectures such as Intel's TrueNorth[18] and the SpiNNaker[19], operate millions of artificial (either hardware or software) neurons whilst smaller neuromorphic approaches run in the 10-100s of thousands[20,21]. These independent successes pave the way towards the networking of such devices into complex and heterogeneous bio-hybrid systems by leveraging the unique advantages of each of the above fields: the capacity to interface to biological neurons, the ability of memristors to act as ultra-compact, tuneable resistive loads (i.e. synapses) and the availability of networks of spiking artificial neurons in on-chip non-von Neumann architectures. An attractive method for achieving this makes use of the standardised interface of the internet, which has thus far been trialled for artificial-to-artificial neural network interfacing only[22,23].

---

[*] Correspondence should be addressed either to SV or TP

We implemented a bNN using three set-ups connected over the internet via UDP (user datagram protocol) comprising a neuromorphic chip hosting artificial neurons (located in Zurich, Switzerland), a memristor handling instrument hosting memristive synapses (Southampton, UK) and a multi-electrode array set-up hosting biological neurons (Padova, Italy), as illustrated in Figure 1. To demonstrate the concept, a simple three neuron network is linked in a feed-forward fashion whereby a pre-synaptic artificial neuron (ANPRE) connects to a biological neuron (BN), which in turn connects to another, post-synaptic, artificial neuron (ANPOST). All connections are realised via memristive synapses (Supplementary Figure 1). Notably, the artificial and biological neuron set-ups communicate exclusively via the synapse set-up (see also Supplementary note 2), thereby creating the two synaptors. The central position of the synapse set-up within the network renders it the *de facto* control centre of the entire system.

The plasticity rule used for the reported benchmarking experiment is a modified, rate-coded Bienenstock-Cooper-Munro (BCM) model[24], where plasticity in synaptors (Long Term Potentiation/Depression: LTP/LTD) occurs upon postsynaptic spiking with the direction of plasticity (i.e. potentiation, depression or no plasticity) determined by the frequency of pre-synaptic activity within a time window leading to post-synaptic spikes[25]. In our modified version, this strictly applies to the reverse biological-to-artificial pathway. In fact, in the forward pathway, plasticity is initiated just by high-frequency pre-synaptic activity, eventually leading to LTP of the synapse[26]. It should be noted that using a rate-coded – and not phase-coded – plasticity rule provides a certain degree of immunity against physical and location-dependent network delays in this experiment, as the specific timing of spikes is secondary in importance to the overall rate. The specific plasticity rule is summarised in Table 1.

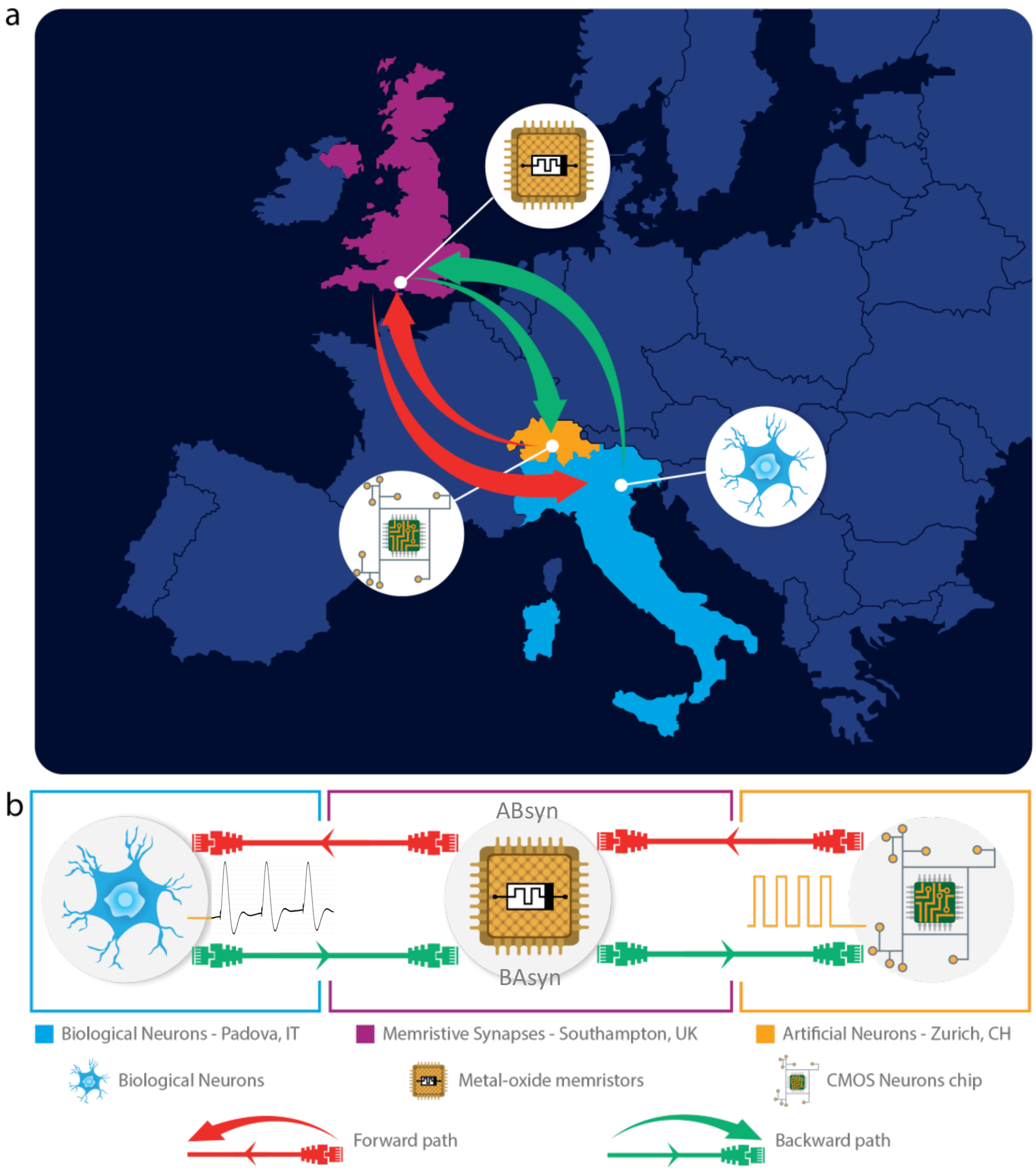

**Figure 1: Geographically distributed bNN set-up. (a) Geographical location of artificial/biological neurons and memristors: biological neurons in Padova (IT), artificial neurons in Zurich (CH) and memristors in Southampton (UK). Forward and reverse signal pathways are illustrated by red and green arrows respectively. (b) High-level connectivity diagram.**

**Table 1: Bienenstock-Cooper-Munro plasticity model parameters used for this work.**

| Pre frequency range (Hz) | BCM plasticity direction |
|---|---|
| <5 | LTD |
| [5, 20] | None |
| >20 | LTP |

Experimental results from the bNN forward path are shown in Figure 2. The input stimulus consisted of artificial neuron ANPRE regularly firing in a forced fashion over four phases: 10, 25, 10 and 4Hz, lasting 20, 20, 20 and 40 seconds respectively. This was designed to induce plasticity changes at the ANPRE to BN synaptor in the direction indicated by the pattern 'none/ LTP/none/LTD', as depicted in Figure 2a. The response of biological neuron BN is recorded by patch clamp and shown in Figure 2b and confirms that only after substantial potentiation has occurred do spikes begin to appear and then remain upon LTP consolidation. Spiking activity persists throughout the subsequent no-plasticity phase and quickly disappears once the LTD induction stage commences. The plasticity underlying this process is also reflected in the amplitude of the excitatory post-synaptic potentials (EPSPs) recorded from the biological neuron (BN), as shown in Supplementary Figure 2. This offers direct evidence on how the weight of the memristor was translated into efficacy of capacitive stimulation used by the cell culture set-up[27], thus effectively demonstrating the implementation of a memristor-weighted, non-invasive synaptor linking an artificial pre-synaptic neuron to a biological post-synaptic neuron. The corresponding memristive synaptic weight evolution is shown in Figure 2c.

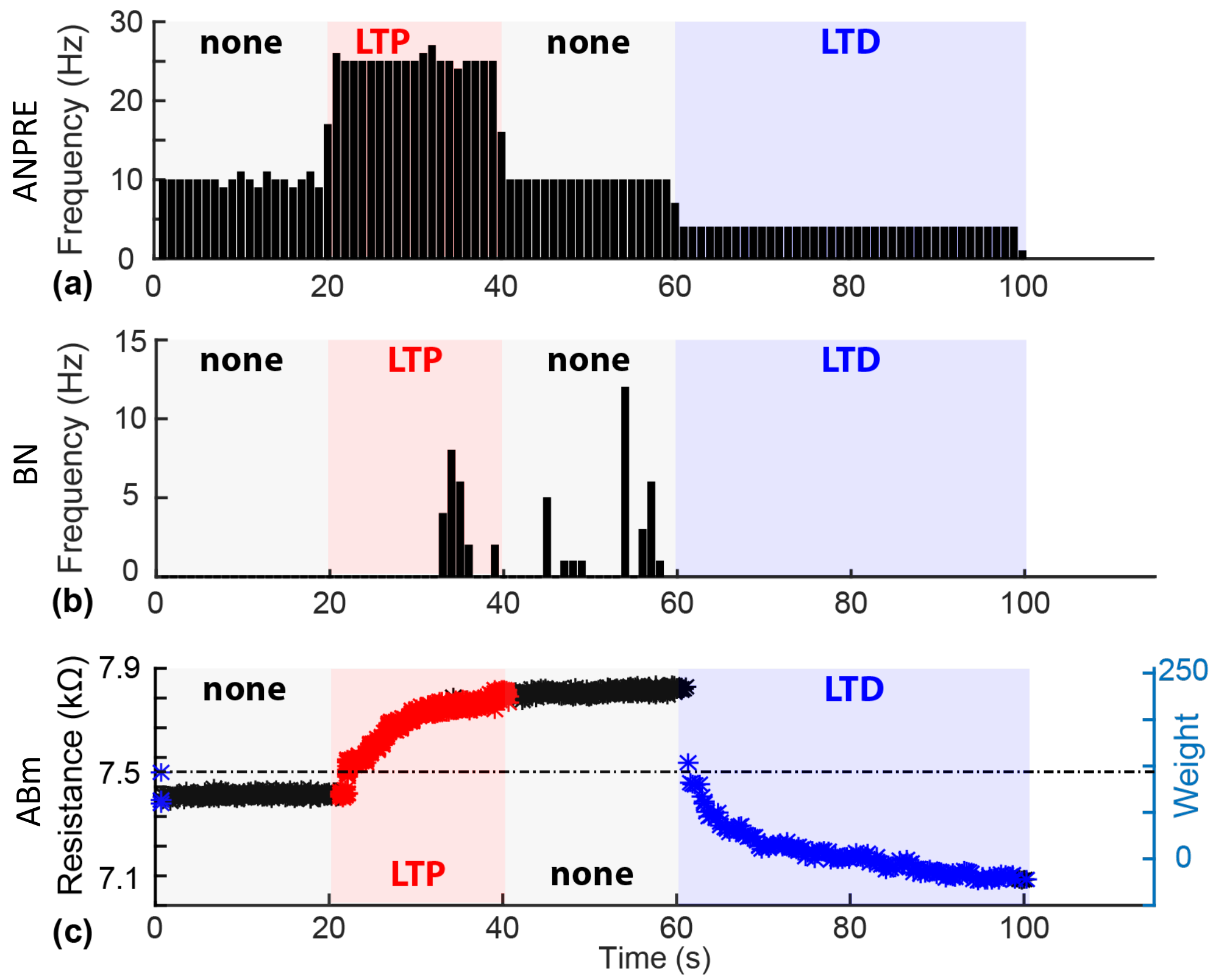

**Figure 2: Geographically distributed bNN experimentally observed operation – forward path. (a) Activity pattern of artificial neuron ANPRE. Firing frequency is modulated in four phases designed to induce plasticity in the synapse linking to the biological neuron (BN) in the direction indicated by the sequence: none/LTP/none/LTD, as per the modified BCM plasticity model. (b) Spike response of biological neuron BN to stimulation from ANPRE. Upon induction, in the late stages of potentiation BN begins responding by firing action potentials (APs). These persist until the commencement of the depression phase. (c) Synaptic weight evolution at the forward (ANPRE to BN) pathway. Each data point denotes a synaptic event that caused LTP (red), LTD (blue) or no further plasticity changes (black). X-axis common to all panels.**

Results from the backward pathway are shown in Figure 3. The spiking activity of BN is forwarded, through memristor BAm, to its post-synaptic target, in this case the artificial neuron ANPOST. ANPOST is set to be spontaneously active in order to guarantee that BCM plasticity will be triggered sufficiently often during the experiment. The effect of spiking at BN then modulates this spontaneous activity. This is evidenced by the increased activity observed in ANPOST coinciding with spiking at BN. This activity dies away as soon as LTD takes place at the forward path, as demonstrated in Figure 3b. The corresponding memristive synaptic weight evolution is shown in Figure 3c and reveals the underlying plasticity. Notably, in contrast to the forward path where tight control of stimulation at ANPRE to BN synapse guarantees reliable induction of LTP and LTD, the persistently low activity at BN leads to either LTD or no plasticity at the BN to ANPOST synapse.

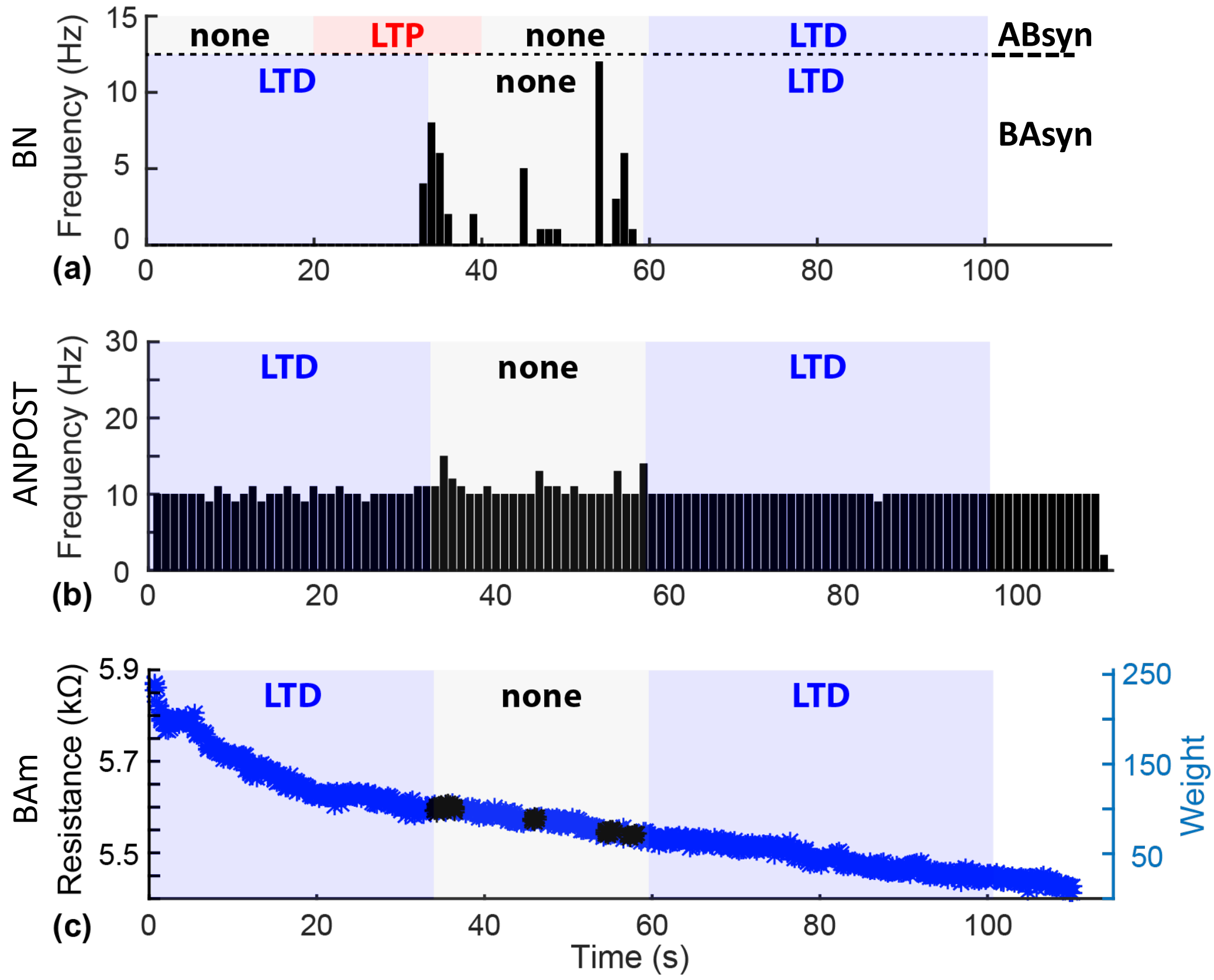

Figure 3: Geographically distributed bNN experimentally observed operation – backward path. (a) Firing response of biological neuron BN to stimulation from ANPRE as reported in Figure 2a. Shadowed areas inidcate plasticity changes occurring at the ABsyn (above the dashed line) and BAsyn (below the deshed line). As previously noted, BN begins responding by firing action potentials during the late stages of potentiation. These persist until the commencement of the LTD induction phase. (b) Spiking frequency measured at artificial neuron ANPOST; the post-synaptic target of BN. Increase in spiking activity is observed in the middle of the run in coincidence with more frequent stimulation from BN followed by a return to baseline, spontaneous activity towards the end. (c) Synaptic weight evolution at the reverse (BN to ANPOST) pathway. Each data point denotes a synaptic event that induced LTP (red), LTD (blue) or no plasticity (black). The rather infrequent activity observed at BN caused the spike rate-dependent plasticity (SRDP) plasticity BCM rule to lead to strong LTD at the reverse pathway synapse and create the LTD/none/LTD plasticity induction pattern shown via blue/gray shadings throughout all panels. LTP/LTD/none plasticity induction window frames used in Fig. 2 are reproduced at the top of 3a for reference. X-axis common to all panels.

Overall, the bNN demonstrates how a feedforward chain of three neurons communicating over long distance is controlled by a single signal input: the forced firing of the neuron at the start of the chain (ANPRE). This creates first a pattern of well-controlled plasticity phases at the forward path synapse and then a less directly controlled pattern of plasticity induction (of the form LTD/none/LTD) at the backward path synapse, as shown in Figure 3c; patterns reproduced in the repeat run are shown in Supplementary Figure 3. Thus, two facts are illustrated: First that the bNN concept and its underlying hardware/software infrastructure operates correctly as evidenced by the forward path. Second, that even for small bNNs (especially if they include biological cells) absolute confidence in the hardware is necessary since interpreting and benchmarking results becomes significantly more difficult the farther away the analysis point moves from the input signal. This is in stark contrast to fully electronic bio-inspired systems, particularly if they are clocked, in which case the nominal operation of the network given some signal input and known initial state can typically be mathematically computed . The introduction of biological cell response and cell state uncertainties (due to the intrinsically changing excitability and the unknown inputs from any other biological neurons in the network) as well as often-unpredictable internet network delays accounts for this loss of determinism.

The bNN in this experiment shows successful operation with artificial and biological neurons interconnected via memristive synapses over the internet. This is achieved by overcoming the issue of UDP propagation delays, which are typically variable and thus difficult to control. To that end, the referencing of secondary partner spikes to the primary partner and use of rate-dependent plasticity both play a vital role. The referencing technique is important in that it makes a remote synapse set-up appear and operate as if it were sitting next to the secondary partner, which implies that if communication from primary to secondary partner is one-way, internet network delays can be de-facto eliminated completely from the operation of the bNN. This one-way referencing is supported by UDP timing measurements we carried out: established connections on a European scale have variable static delays from 10-90ms, but timing of individual UDP packets along a connection varies below 2ms, i.e. the relative timing of pulses is stable. However, completely compensating for round-trip delays may only be possible for very special cases, i.e. where relative spike delay is relevant, but spike delay between neuron populations (at different geographic locations) can be large.

Despite these limitations, our bNN represents the first example in nature of a geographically distributed hybrid network of artificial and biological neurons remotely connected through synaptic-like elements. Whereas brain evolution had to face tight physical constraints of spatially confined connectivity and limited conduction velocity through neuronal branches, the proposed bNN is a first embryonic example of networks where such barriers are overcome and that could globally process real-world inputs in a mixed biological/bio-inspired fashion. Synaptors are the major building blocks of these systems, allowing for plasticity-driven communication and processing of signals at the edge of hybrid links. In perspective, this bNN can be seen as the first step for conceptually novel neuroprostheses and electroceuticals. Synaptic-like memristive connections between artificial and biological neurons will enable the use of neuromorphic devices and architectures as surrogates of native neuronal circuits, assisting repair and rehabilitation in neurological disorders though smart implantable neural interfaces. Simultaneously, in this work we presented the necessary software and hardware ingredients to create a highly scalable 'Internet of Neuroelectronics', with lab setups all over the world able to participate in biohybrid experiments no single lab would be capable of. Our approach mitigates scaling constraints faced by all individual platform technologies (it's easy to scale up highly optimised platforms for cell culturing, memristive arrays, etc), although scaling will obviously necessitate a corresponding bandwidth increase.

## Methods

The central part of the artificial side of the bio-hybrid system is formed by a reconfigurable on-line learning spiking neuromorphic processor (ROLLS)[20], which contains neuromorphic CMOS circuits emulating short-term plasticity (STP) properties of synapses[28] and long-term plasticity (LTP) ones[29]. In addition, this processor comprises mixed signal analogue-digital circuits which implement a model of the adaptive exponential integrate-and-fire neuron[30]. Input and output spikes are sent/transmitted from the chip using asynchronous IO logic circuits which employ the Address-Event-Representation (AER) communication protocol[31]. The chip is connected to a host PC which receives UDP-packets from the internet. These packets contain information on stimulus destinations and corresponding synaptic weights. This information is decoded by a Field Programmable Gate Array (FPGA) device and conveyed to the neuromorphic processor. In this work, the parameters of the CMOS synapse circuits were set to produce weak excitatory postsynaptic currents (EPSCs) with long time constants, such that high frequency stimulation causes an additive effect on the net amplitude of the resulting EPSC. The value of the weight encoded in the UDP packet was used to produce spike trains of different frequencies transmitted by the FPGA to the neuromorphic processor (see also Supplementary Figure 4). In addition to the signals arriving from the UDP interface, locally generated spike trains were sent to the neuromorphic processor, to provide a controlled stimulus for evoking background activity. This system is shown in Supplementary Figure 4.

The memristive synapse set-up consists of an array of metal-oxide devices positioned inside a commercially available instrumentation board[17] (Supplementary Figure 5). The instrument is controlled by a PC, which handles all the communications over UDP. During operation, arriving UDP packets carry information related to the identity and firing timing of either artificial or biological neurons. Once a packet is received, the neural connectivity matrix is consulted in order to determine which neurons are pre- and which are post-synaptic to the firing cell. Then, if the plasticity conditions are met, the instrumentation board applies programming pulses that cause the memristive synapses to alter their resistive states. The set-up has control over whether LTP- or LTD-type plasticity is triggered in each case, but once the pulses have been applied it is the device responses that determine the instantaneous learning rate. Notably, resistivity transitions of the device are non-volatile, they hold over at least hours[32] as also exemplified in our prototype experiment and are therefore fully compatible with typical LTP and LTD time scales of natural synapses. Because of its central role in the network, the memristive synapse set-up holds both the connectivity matrix defining the network connectome and the record of all spiking activity in addition to controlling the handling of time.

Embryonic (E18) rat hippocampal neurons were plated and cultured on the MEA according to procedures described in detail in[33]. Recordings were performed on 8-12 DIV neurons. The experimental setup in UNIPD (Supplementary Figure 6a,b & c) enables UDP-triggered capacitive stimulation of neurons[34] while simultaneously recording and communicating via UDP the occurrence of action potentials (APs) detected by the patch clamp electrode. The MEA is controlled by a dedicated stimulation board and all the connections to partners, Southampton and Zurich, are managed by a PC running a LabVIEW-based software. The stimulation of the neurons is operated through a planar CMOS Multi Electrode Array (MEA) with 20x20 independent $TiO_2/ZrO_2$ capacitors, each one with an area of (50 x 50 $\mu m^2$). The stimulation protocol has been developed according to an approach described in detail in[34] and further optimized for non-invasive tuneable stimulation of the cell. In brief, capacitive stimulation allows for tuning neuronal excitation within the subthreshold range and up to the firing of the action potential by varying the repetition number of appropriate stimulation waveforms (Supplementary Figure 6d). In this way, by converting the memristor resistance in stimulation cycles through the capacitive microelectrode, a synaptic-like connection is created between artificial and biological neuron, which translates presynaptic firing into subthreshold responses (i.e. approximating EPSPs) or action potentials depending on synaptic weight.

Patch-Clamp recordings were performed in whole-cell current-clamp configuration using an Axopatch 200B amplifier (Molecular Devices, USA) connected to the PC through a BNC-2110 Shielded Connector Block (National Instruments Corp, Austin, TX, USA) along with a PCI-6259 PCI Card (National Instruments Corp, Austin, TX, USA). WinWCP (Strathclyde Electrophysiology Software, University of Strathclyde, Glasgow, UK) was used for data acquisition. Micropipettes were pulled from borosilicate glass capillaries (GB150T-10, Science Products GmbH, Hofheim, Germany) using a P-97 Flaming/Brown Micropipette Puller (Sutter Instruments Corp., Novato, CA, USA). Intracellular pipette solution and extracellular solution used during the experiments were respectively (in mM): 6.0 KCl, 120 K gluconate, 10 HEPES, 3.0 EGTA, 5 MgATP, 20 Sucrose (adjusted to pH 7.3 with 1N KOH); 135.0 NaCl, 5.4 KCl, 1.0 MgCl2, 1.8 CaCl2, 10.0 Glucose, 5.0 HEPES (adjusted to pH 7.4 with 1N NaOH). Digitised recordings were analysed by a custom LabVIEW software running on the PC, allowing detection and discrimination of firing and EPSP activity through a thresholding approach.

**Author contributions**

The experiments were jointly conceived by TP, SV and GI, who share senior authorship. The experiments were jointly designed and ran by AS, AC, RG, who are acknowledged as shared first authors. AK manufactured the memristive devices. FR and MR assisted with the biological system set-up and operation. CM provided valuable feedback and guidance during the write-up of the paper. The paper was jointly written by all co-authors.

# SUPPLEMENTARY MATERIAL

## Supplementary figures

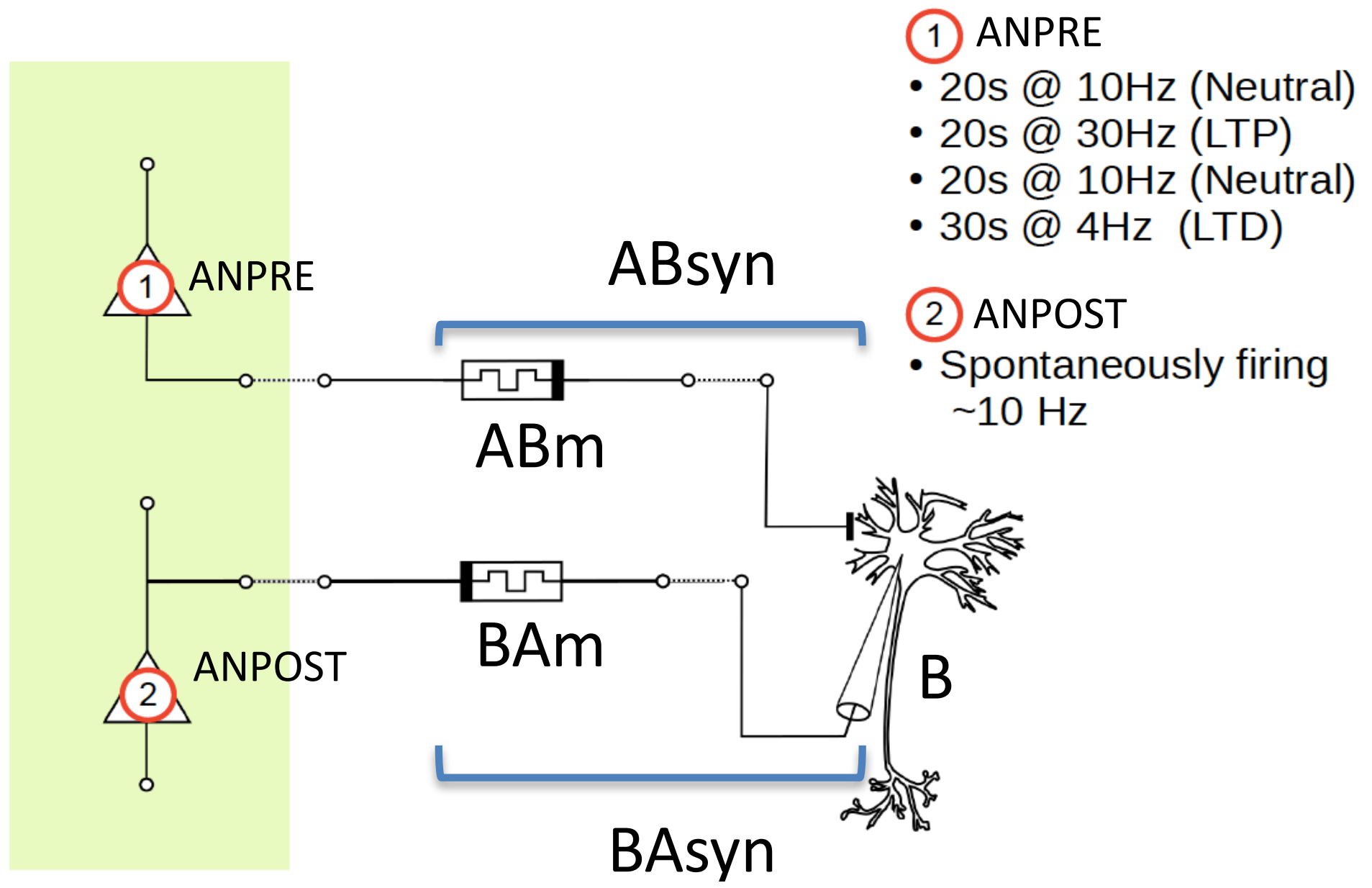

Supplementary figure 1: bNN layout, with two memristors (BAm and ABm) establishing the two synaptors: ABsyn (Forward path with **ANPRE** presynaptic to **BN**) and BAsyn (Backward path, with **BN** presynaptic to A**NPOST**). ABsyn: ANPRE spikes are fed to ABm whose resistivity changes tune extracellular stimulation of BN through a capacitive microelectrode array (see Methods and supplementary Figure 6). BAsyn: BN activity (sub- and supra-threshold) is monitored by a patch-clamp pipette. Spikes are fed to BAm which drives ANPOST firing by modulating the EPSC (see Supplementary Figure 4).

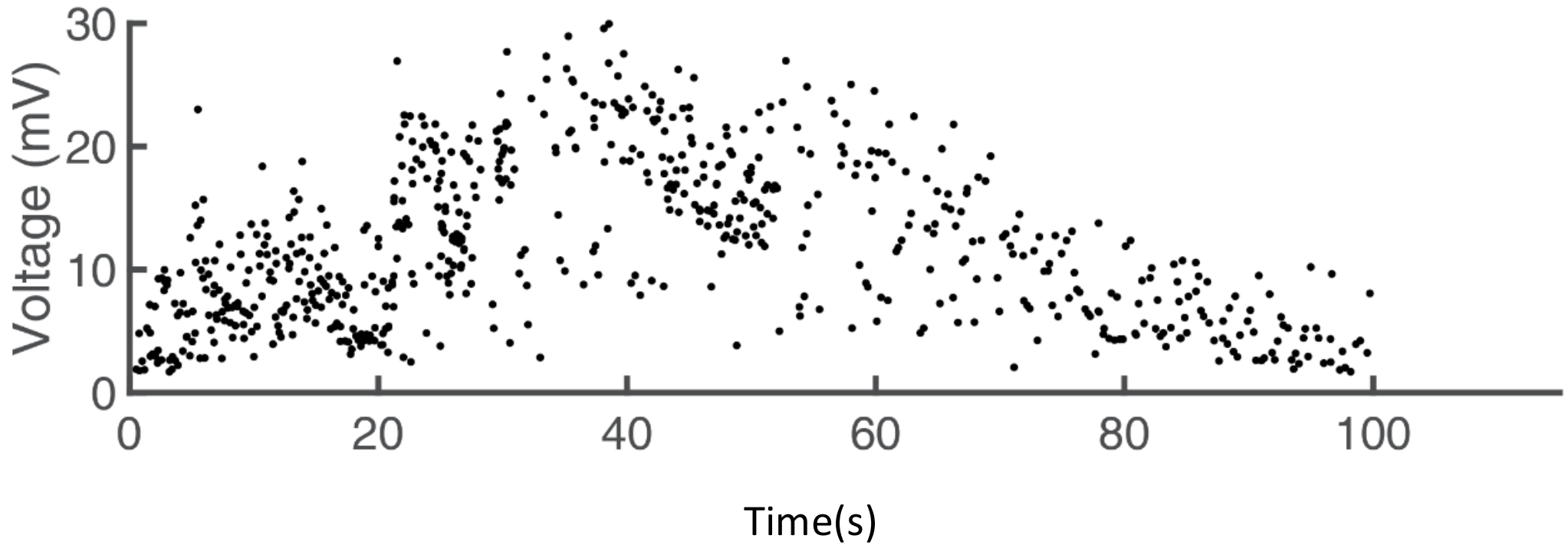

**Supplementary figure 2: amplitude of subthreshold events measured at BN in response to spikes arriving via UDP from ANPRE. Amplitude modulation is observed which is dependent on ABm memristive synaptic weight and consequent tuning of capacitive stimulation of the neuron (see Supplementary Figure 6).**

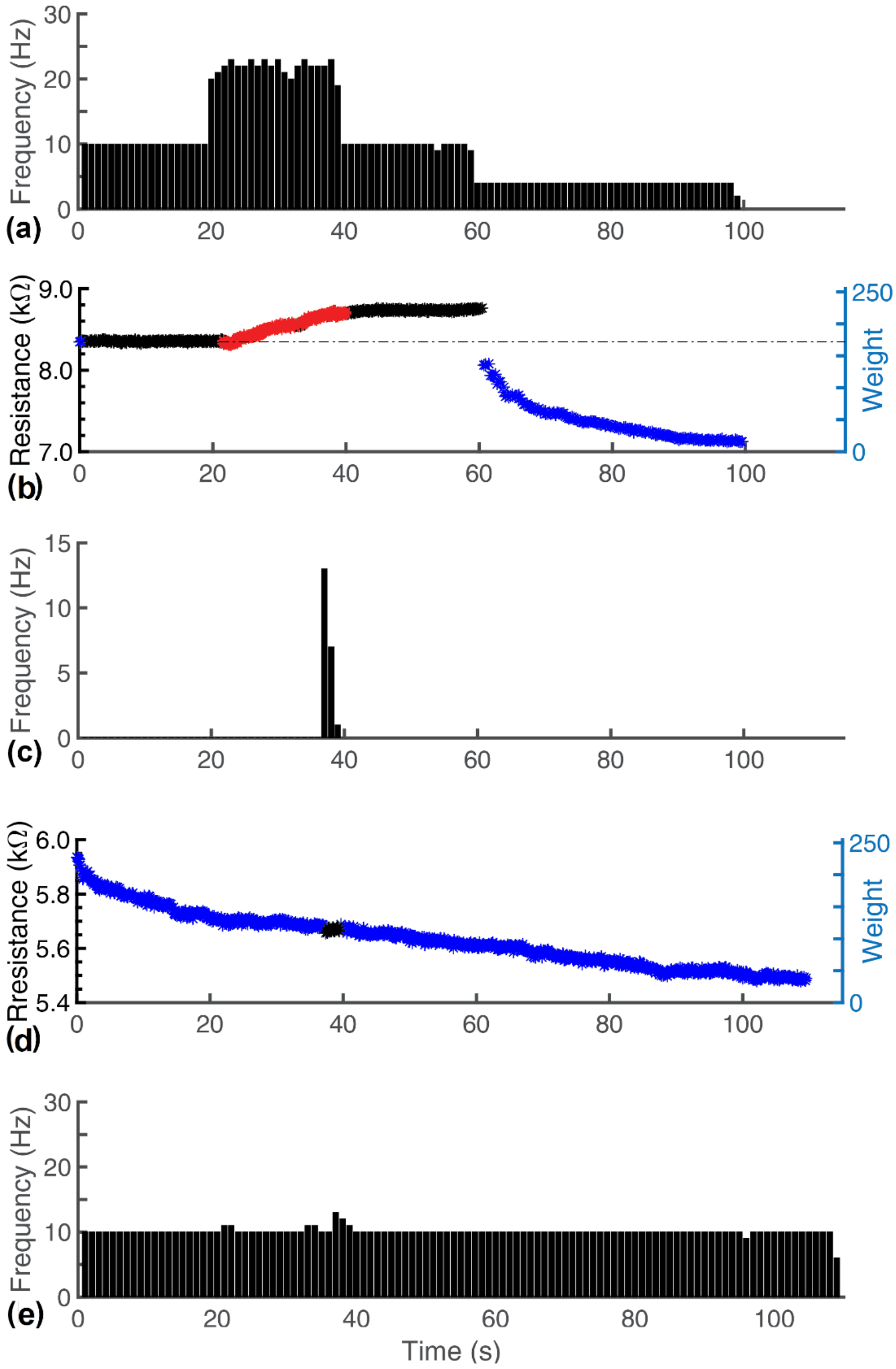

**Supplementary figure 3: repetition of the experiment shown in Figures 2 and 3. (a) ANPRE activity pattern. (b) ANPRE to BN plasticity evolution. (c) BN activity pattern. (d) BN to ANPOST plasticity evolution. (e) ANPOST activity pattern. Differences in firing rates with respect to experiment in Figures 2 and 3 are due to different initial synaptic weights, weight evolutions and biological neuron responses. However, the overall response remained qualitatively the same, proving both reliability and flexibility of the system to complex dynamics.**

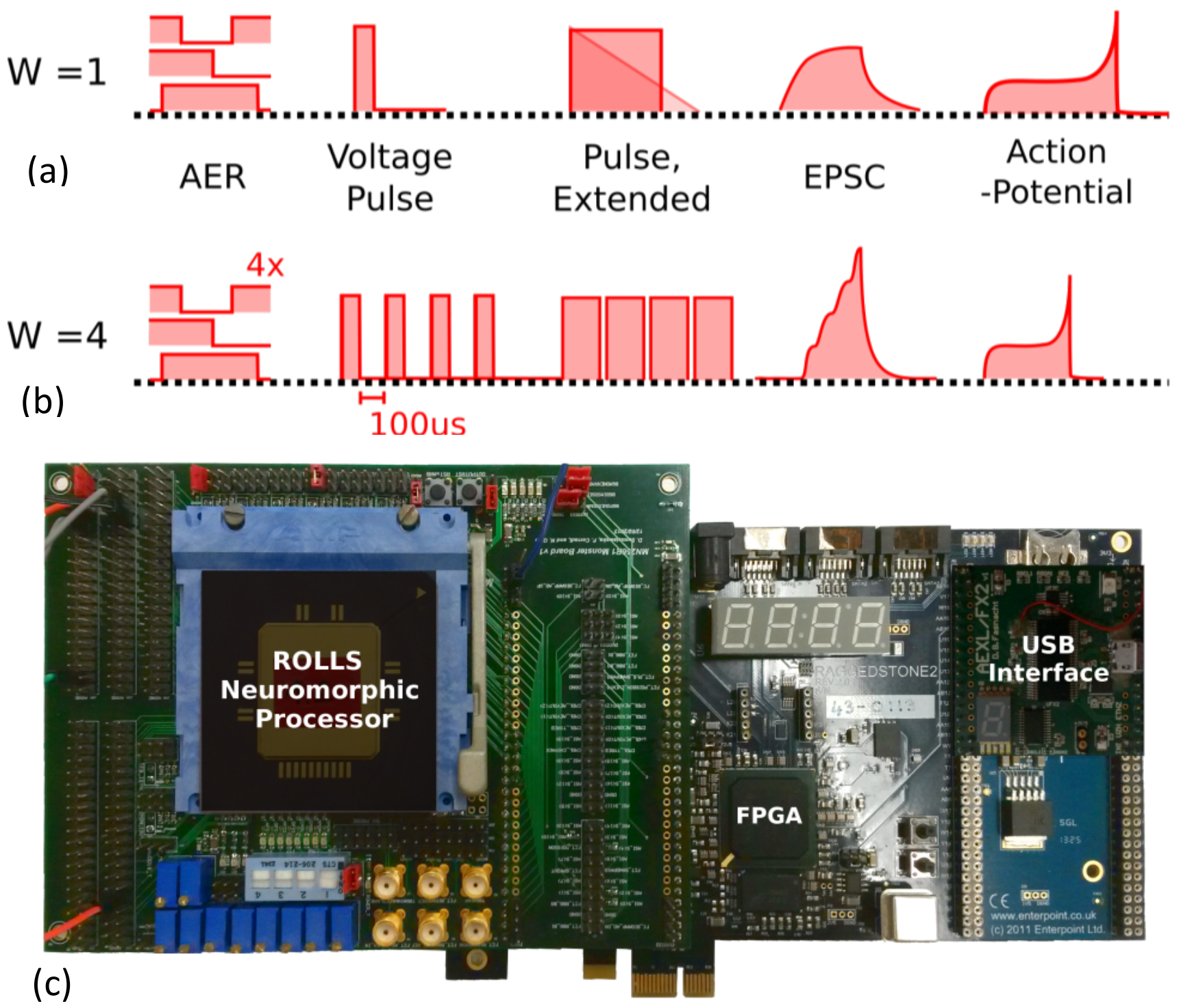

**Supplementary figure 4: Artificial neuron set-up, located in Zurich, Switzerland. (a,b) Two examples of how an incoming UDP packet is handled in the cases where the input weight is equal to 1 – i.e. minimum weight quantum – (a) and 4 respectively (b). Note the difference in the EPSC waveforms generated as a result. (c) Artificial neuron set-up located in Zurich, Switzerland.**

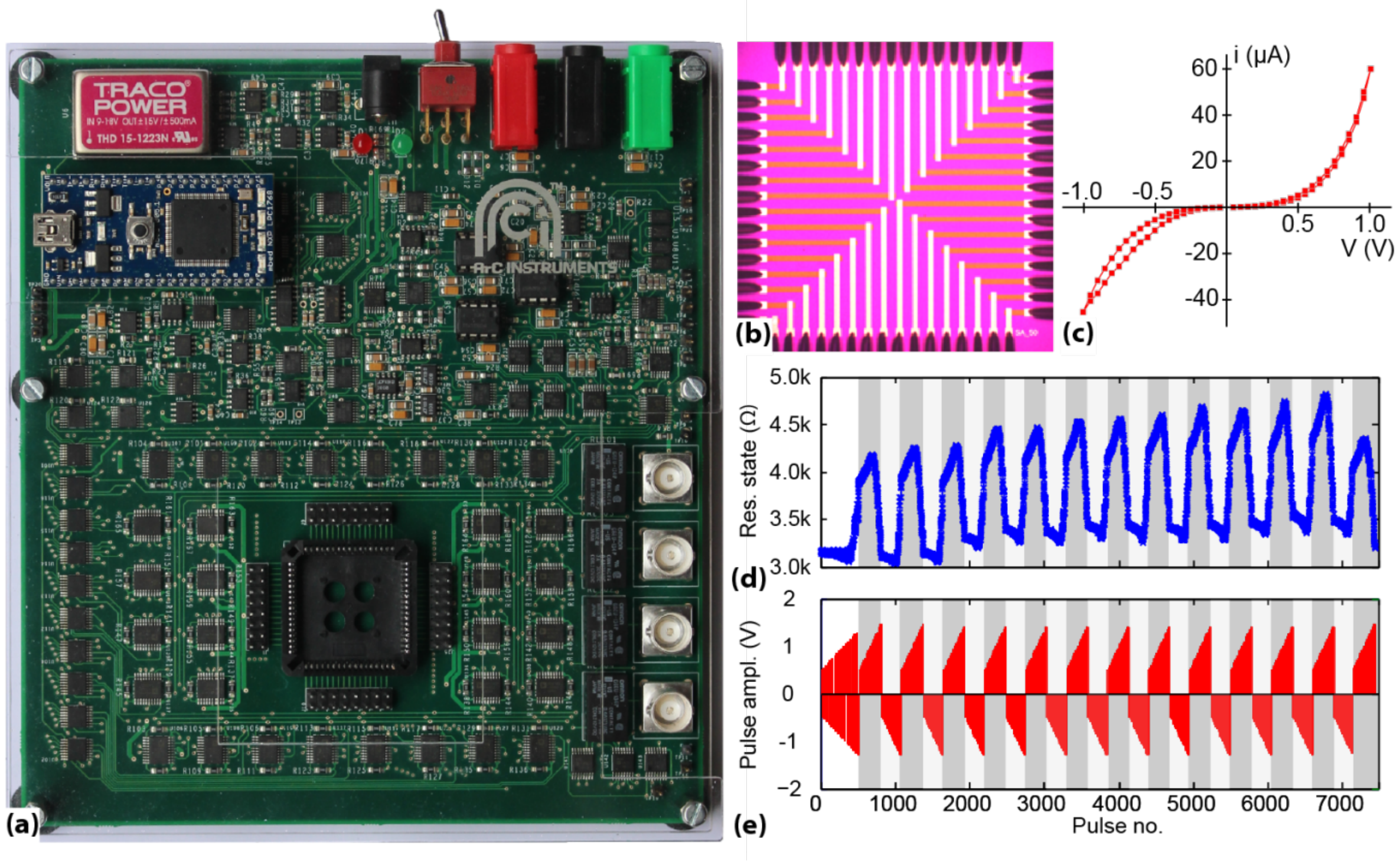

**Supplementary figure 5: Memristive synapse set-up core located in Southampton, UK. (a) ArC board. The devices sit within the PLCC68 package socket featuring prominently at the lower middle of the board. (b) Example of stand-alone device array (microphotograph). (c) Typical current-voltage (i-V) characteristic of metal-oxide AlOx-TiOx bilayer devices used in this work. A small amount of pinched hysteresis is observable in this deliberately low voltage/low invasiveness i-V run. (d) Device resistive state modulation in response to stimulation in (e). This data is taken from[35].**

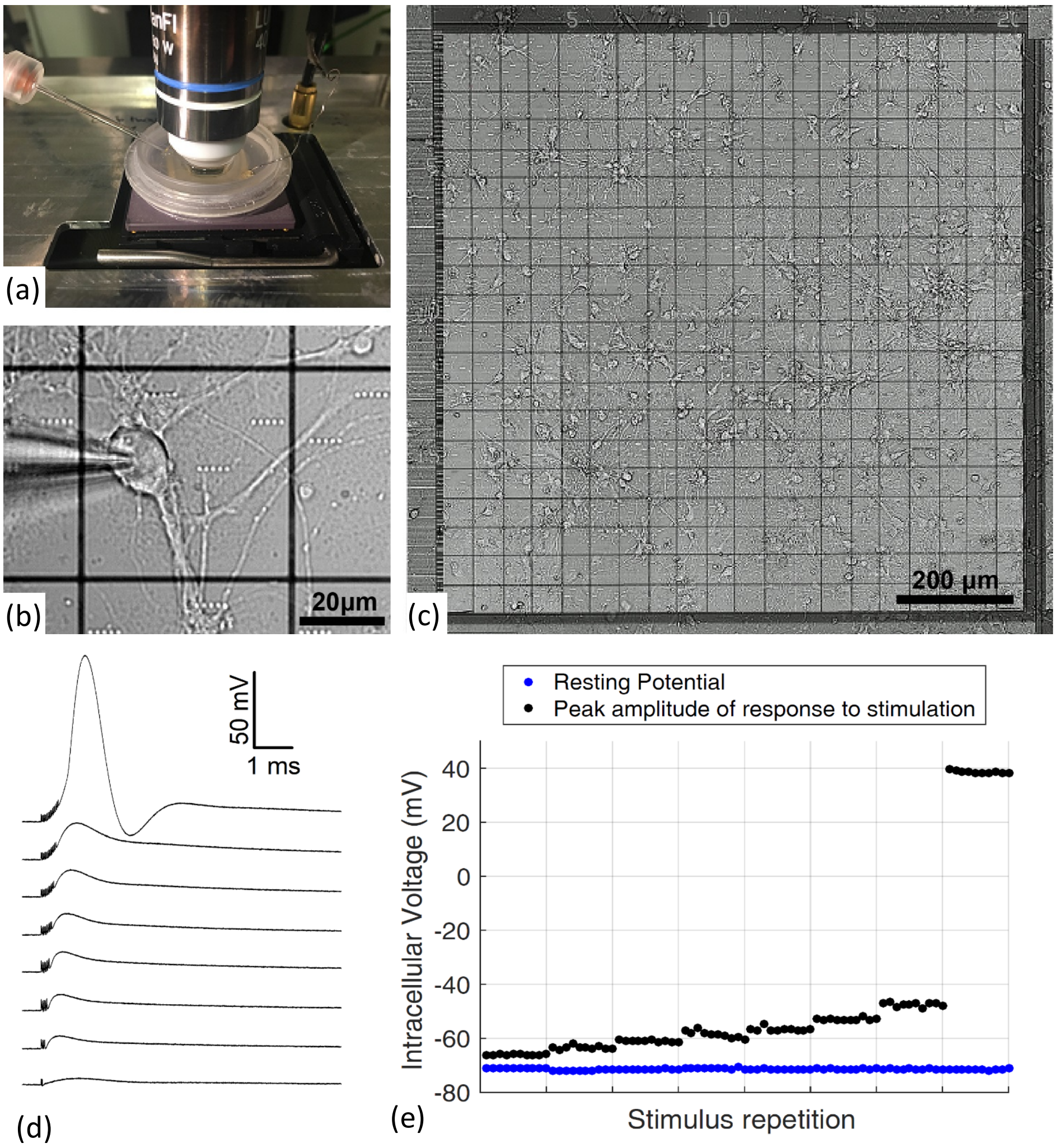

**Supplementary figure 6: Biological neuron set-up, located in Padova, Italy. (a) MEA chip mounted on a stimulation board. A Perspex chamber on the chip contains the culture medium or the extracellular solution for the experiment. A patch pipette is seen on the left. (b) Patch pipette recording a neuron on one of the square capacitive stimulation spots of the 20 x 20 array. (c). (d) Examples of whole cell patch clamp recordings showing the neuron response to extracellular capacitive stimulation. Eight different stimulation protocols were applied with increasing number of pulse repetitions (from bottom to top: 2,4,6,8,10,12,14,16). Subthreshold responses with increasing amplitude and resembling EPSPs are evoked when applying 2 to 14 pulses. Threshold for AP firing is reached at 16 pulses. (e) Peak intracellular voltage values of neuronal responses when repeating each protocol ten times (1s interval) in the same neuron (black). Cell resting potential just before stimulation (blue).**

| R1 | Neuron ID | R2 | Timestamp |
|---|---|---|---|
| 8 bits | 24 bits | 8 bits | 24 bits |

**Supplementary figure 7: UDP packet structure used in this work. The packet consists of 64 bits in total, broken into four segments as indicated by the labels. Labels correspond to standard use of packets in AER protocol with segments R1 and R2 corresponding to flexible space to be used by various applications utilising the AER packet format. The specific uses of each segment in this work are summarised in Supplementary table 1.**

## Supplementary tables

| | R1 | Neuron ID | R2 | Timestamp |
|---|---|---|---|---|
| **Padova (secondary)** | Partner identifier (primary/secondary) | Spiking neuron ID | Type of event (PSP, forced AP, spontaneous AP) | Absolute time |
| **Southampton (synapse)** | Partner identifier (primary/secondary) | Postsynaptic neuron ID | Weight | Absolute time |
| **Zurich (primary)** | Partner identifier (primary/secondary) | Spiking neuron ID | -nothing- | General relative time |

**Supplementary table 1: UDP packet payloads by packet segment and partner. Cell entries refer to the type of packet produced by each partner (not the type of packet received by each partner).**

## Supplementary notes

### Supplementary note 1: UDP packet structure and contents.

The UDP packets used in this work follow the structure shown in Supplementary Figure 7 and carry a total of four variable values as payload. The meanings of these variables for each partner are summarised in Supplementary Table 1. For all partners, segment R1 contains a partner identifier value specifying whether the packet is arriving from a primary, synapse or secondary partner. The neuron ID segment contains the identity of the neuron firing if this is emitted by a partner hosting neurons (primary/secondary). This is all the information required by the synapse partner in order to compute which post-synaptic neurons need to be stimulated through the memristor-based synapses. As a result, the synapse partner sends the ID of the postsynaptic neurons to be stimulated. Note that whist the primary and secondary partners send a packet per neuron firing, the synapse partner then emits a packet for each stimulated synapse in response. Segment R2 is used by the biological neuron partner to inform the synapse partner on whether the event that is being communicated in each packet is a PSP, an action potential (AP) resulting from stimulation or a spontaneous AP. The synapse partner uses the same segment to communicate the strength of the stimulation corresponding to each stimulated synapse, in effect the weight of the synapse. Finally, the timestamp segment communicates timing information using different protocols for each partner. Timing management details are covered in Supplementary note 2.

### Supplementary note 2: Timing protocol summary.

The synapse set-up, being the node that links biological and artificial partners together, controls the overall handling of time during operation. Under this system, one of the partners (in this case Zurich, which hosts the neuron that starts the signal path - ANPRE) is labelled as the 'primary partner' and all timing information arriving from that partner is treated as ground truth. Spiking activity from the secondary partner (or in the case of larger systems: partners) is then referenced against this forced ground truth. For example, if the primary partner asserts that neuron X fires a spike at time $t_0$, then the secondary partner is informed of this, and responds in the form of a post-synaptic potential (PSP) evoked at $t_0$ through the synapse set-up. If then a neuron Y in the secondary partner set-up fires $\Delta t$ (wall-clock) time units after being informed of the firing of neuron X, it emits a packet informing the synapse set-up that e.g. neuron Y fired at time $t_0 + \Delta t$. This way the relative timing between spikes arriving from the primary partner and the spikes triggered in response, by the secondary partner, is maintained despite any network delays. In exchange, when the secondary partner wishes to communicate spikes to the primary partner, all network delays for the entire round-trip will burden that communication.

Our design ensures that at least in the pathway from primary to secondary partner, timing control is sufficiently tight to sustain timing-sensitive plasticity.

Each partner in the set-up handles time differently: The primary partner operates in 'general relative time', whereby UDP packets inform the synapse set-up of the identity of the neuron currently spiking and the time interval between the present spike and the previous spike, regardless of the origin of the previous spike. Therefore, if neuron 1 spikes at time 12000 and neuron 2 spikes at time 12012, the UDP packet will contain the information: ID=1, dt=12. The synapse set-up operates on the basis of absolute time, i.e. it constructs a sequence of events on a constantly advancing time axis and translates general relative timestamps arriving from the primary partner into absolute time. It then informs all partners of the spiking of every other partner in absolute time. Finally, the secondary partner operates on the basis of a wall-clock that is reset every time information on a primary partner spike arrives (relative time). The secondary partner communicates its own spikes back to the synapse set-up in absolute time.